\documentclass{osa-article}
\usepackage{bm}
\usepackage{braket}
\journal{oe}

\articletype{Research Article}

\begin{document}
\title{Valley-dependent Corner States in Honeycomb Photonic Crystal without Inversion Symmetry}
\author{Huyen Thanh Phan,\authormark{1,*} Feng Liu,\authormark{2,3,4,**} Katsunori Wakabayashi \authormark{1,5,6,***}}
\address{\authormark{1}Department of Nanotechnology for Sustainable
Energy, School of Science and Technology, \\
Kwansei Gakuin University, Gakuen 2-1, Sanda, Hyogo 669-1337, Japan\\ 
\authormark{2} School of Physical Science and Technology, Ningbo University, Ningbo,
315-211, China\\
\authormark{3} Institute of High Pressure Physics, Ningbo University, Ningbo, 315-211, China\\
\authormark{4} Laboratory of Clean Energy Storage and Conversion, Ningbo University, Ningbo, 315-211, China
\authormark{5}Center for Spintronics Research Network (CSRN), Osaka University,
Toyonaka 560-8531, Japan\\
\authormark{6}National Institute for Materials Science (NIMS), Namiki 1-1,
Tsukuba 305-0044, Japan}

\email{\authormark{*}phanthanhhuyenpth@gmail.com}
\email{\authormark{**}liufeng@nbu.edu.cn}
\email{\authormark{***}waka@kwansei.ac.jp} 


\begin{abstract}
 We study topological states of honeycomb photonic crystals in absence
 of inversion symmetry using plane wave expansion and finite element methods.
 The breaking of inversion symmetry in honeycomb lattice leads to
 contrasting topological valley indices,
 i.e., the valley-dependent Chern numbers in momentum space.
We find that the topological corner states appear for
 60$^\circ$ degree corners, but absent for other corners, which can be
 understood as the sign flip of valley Chern number at the corner.
Our results provide an experimentally feasible platform for exploring
 valley-dependent higher-order topology in photonic systems.
\end{abstract}

\section{Introduction}

The application of mathematical concept, {\it topology}, to solid-state
physics~\cite{TopoBandTheo, TopoInsu1,TopoInsu2,Ando2013} 
has resulted in fruitful
achievements to design functional quantum materials and
devices~\cite{TopoInsu4, TopoInsu8, TopoInsu5, TopoInsu3, TopoInsu6,
Kim2019, Okuyama2019, Kong2010}. 
For example, in topological insulators, owing to the nontrivial topological invariant of
bulk wave functions, topologically protected edge states appear and provide robust electronic conduction even in the presence of local perturbations such as defects and edge roughness.
Those topologically protected surface states are promising candidates for ultra low-power-consumption
electronics and quantum computation~\cite{He2019}. 
Furthermore, not only the topologically protected edge states, the
topologically protected corner states were recently found as a
consequence of higher-order topology of bulk wavefunction~\cite{Ezawa2018, Ezawa2018a, Luo2019}.
Inspired from the topological materials, the concept of topology
has been extended to the photonic crystals (PhCs) and intensively studied~\cite{Wang2020a, Lu2014, Ozawa2019, Ota2020}.
PhCs are artificially periodic structures that consist of dielectric
media and meta-materials, where the electro-magnetic (EM) waves are
described by Bloch functions similar to electron waves in crystalline
materials. As analogies of insulators, dielectric PhCs can also prevent EM wave of
certain frequencies from propagating through the periodic structure,
resulting in the formation of photonic energy band gaps. 

Since it is demanded to design and fabricate the photonic devices to
provide lossless EM flow for the application of optical telecommunication
and computing, studies of topological PhCs have drawn
much attention and grown rapidly. Shortly after the theoretical
proposal~\cite{Haldane2008}, 
the topologically protected unidirectional propagation of EM waves has been experimentally observed using two-dimensional (2D) magneto-optical PhCs in microwave regime~\cite{Wang2008, Wang2009}. The research has stimulated further theoretical works on the PhCs waveguides~\cite{Hafezi2011, khanikaev2013,Skirlo2014} and also experimental studies on topological edges states in PhCs ~\cite{Exp1, Exp2}. Besides,
the photonic analogy of topological electronic materials has been
realized such as the photonic Chern insulator~\cite{Chi2,Chi6}
as a result of the broken time-reversal symmetry. In this
system, Berry curvature has non-zero distribution in the momentum space, leading
to non-zero Chern number after integrating Berry curvature over the momentum
space. Moreover, even in the absence of Berry curvature, topological properties of PhCs can still be found in
2D square PhCs~\cite{0BC,SquareOta,2ndorderTopo,Chen2019}.
The photonic band structures of zero Berry curvature systems
show the topologically non-trivial band gaps
and the localization of EM wave at one-dimensional (1D) edges and 
zero-dimensional (0D) corners were also observed~\cite{SquareOta,
2ndorderTopo}. 
These topological states are explained by
vectored Zak phase and gapped Wannier bands~\cite{Delplace2011, Liu2019,
Benalcazar2017, Xue2019}, unlike Chern number in previous 
cases. Not only Chern number and Zak phase, the internal degree of
freedoms like valley indices can also be used to characterize
topological properties. Valley means the local minima or maxima of
energy band structures. For example, graphene, which provides honeycomb
network of $\pi$-electrons, has such valley structures
at Dirac points in momentum space, which appear at non-equivalent $K$
and $K^\prime$ points in the 1st BZ. 
In graphene, these valley indices work as pseudo-spin degree of freedom
and can be detected through valley Hall
effect~\cite{Xiao2007}. In analogy to graphene, valley-Hall effect can
also be explored in photonic systems~\cite{Zhen2017,JihoNoh2018,Zhang2018,Lu2018,Ni2018}, where it can explain for not only
the first order topological states but also higher-order topology in
photonic systems.

In this paper, we theoretically study the topological states of 2D
honeycomb PhC without inversion symmetry, where
two kinds of dielectric rods are placed to form the honeycomb lattice in air.
Owing to the broken inversion symmetry, the system has non-equivalent valley degrees of
freedom in momentum space.
Therefore, Berry curvatures have non-zero distribution in momentum space, leading to the non-trivial
topological indices~\cite{BC1,BC2,BC3,BC4}. 
The appearance of topological corner states as higher order topology are the results of interactions 
between non-equivalent
valleys in honeycomb PhC. Our results suggest a promising
platform of future experimental studies to explore valley dependence of higher-order topology in photonic systems and provide the
possibility to the application of topological PhCs in optical communication using valley degree of freedoms.

\section{2D Honeycomb PhC with broken inversion symmetry and valley Chern number} 
\begin{figure}[ht!]
\centering\includegraphics[width=0.8\textwidth]{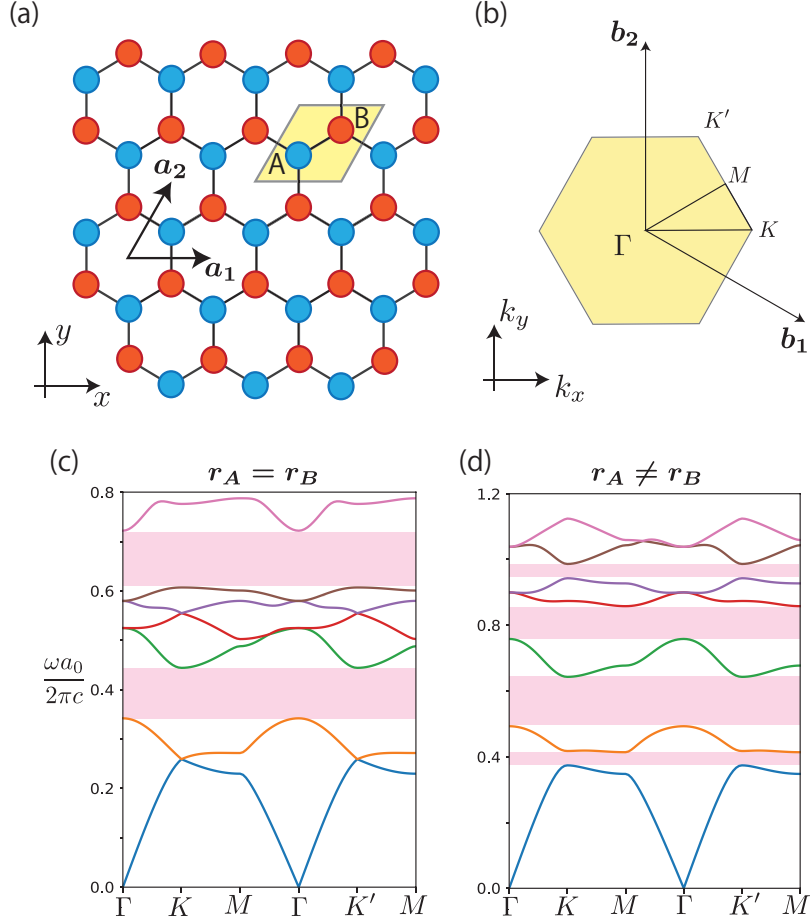}
\caption{(a) Schematic of 2D honeycomb PhC. $\boldsymbol{a_1}=a_0(1,0)$ and
 $\boldsymbol{a_2}=a_0(1/2,\sqrt{3}/2)$ are primitive vectors. Here,
 $a_0$ is the lattice constant. Solid yellow rhombus indicates the
 primitive unit cell, which contains two non-equivalent dielectric rods
 called A and B, colored with cyan and magenta, respectively. The radii for these rods are $r_A$ and $r_B$,
 respectively.  (b) The corresponding 1st BZ 
 of 2D honeycomb lattice with high symmetric points. $\bm{b_1}$ and $\bm{b_2}$ are primitive
 reciprocal lattice vectors.
 (c) Photonic band structure for 2D honeycomb PhC
 which converses inversion symmetry ($r_A=r_B=0.18a_0$), showing the Dirac points at $K$ and
 $K^\prime$ point. (d) Photonic band structure for PhC with broken inversion symmetry
 $r_A=0.0825a_0$ and $r_B=0.1a_0$.} 
\label{figure1}
\end{figure}

Here we briefly give an overview of photonic states in 2D honeycomb PhC with broken inversion symmetry and introduce the valley Chern number. 
Figure~\ref{figure1}(a) shows the lattice structure of 2D honeycomb PhC,
where the solid rhombus indicates the unit cell. There are two
non-equivalent lattice sites called A and B in the unit cell, where the
radii for cylinders A and B are defined as $r_A$ and $r_B$,
respectively. Throughout this paper, we assume that the dielectric
cylinders are Yttrium Iron Garnet (YIG) with dielectric constant
$\varepsilon = 15$, and those are arranged periodically in honeycomb
lattice with the period of $a_0$ in air. 
The eigenvalue equation for EM wave in periodic media is given as follow:
\begin{equation}
    \frac{1}{\varepsilon \left(\boldsymbol{r} \right)}
     \,\boldsymbol{\nabla} \times \boldsymbol{\nabla} \times
     \boldsymbol{E}\left(\boldsymbol{r}\right) =
     \frac{\omega^2}{c^2}\boldsymbol{E}\left(\boldsymbol{r}\right),
\end{equation}
where $\boldsymbol{E}\left(\boldsymbol{r}\right)$ is vector electric
field and $\varepsilon \left(\boldsymbol{r} \right)$ is dielectric
function. $\bm{r}=(x,y)$ is the position vector in 2D space. We assume that
the perpendicular direction to the 2D plane is translational invariant.
$c$ is the speed of light. $\omega$ is the eigenfrequency.
Figure~\ref{figure1}(b) shows the corresponding 1st Brillouin Zone (BZ) for 2D
honeycomb PhC.

In Fig.~\ref{figure1}(c), photonic band structure is shown for
$r_A=r_B=0.18a_0$. In this case, the system preserves inversion symmetry
and becomes the photonic analogue of graphene.
We calculate the photonic band structure by using the plane wave expansion method~\cite{joannopoulos2008molding}.
Here the unit of angular frequency is $2\pi c/a_0$.
As we can see, the linear Dirac dispersion appears at $K$ and $K^\prime$
points. In this sense, the present system can be considered as the photonic analog of graphene. 
However, the complete photonic band gaps are opened around the frequencies of $0.4$
and $0.65$, which are intrinsic feature of 2D honeycomb PhCs.

Next we shall consider the case of broken inversion symmetry by making two rods
different radii, i.e., $r_A\neq r_B$. 
Figure~\ref{figure1}(d) shows the photonic band structure with broken
inversion symmetry, i.e., $r_A=0.0825 a_0$ and $r_B=0.1 a_0$. 
Compared with the photonic band structure in
Fig.~\ref{figure1}(c), the degeneracies at Dirac points $K$ and $K^\prime$ are lifted, leading to the complete band gaps with parabolic
dispersion. This is the photonic analogue of electronic states for
hexagonal Boron-Nitride ($h$BN).
In this paper, we mainly focus on the 1st lowest photonic band gap,
which originates from lifting the degeneracy at Dirac
points between the 1st and the 2nd lowest photonic bands. 

\begin{figure}[ht!]
\centering\includegraphics[width=\textwidth]{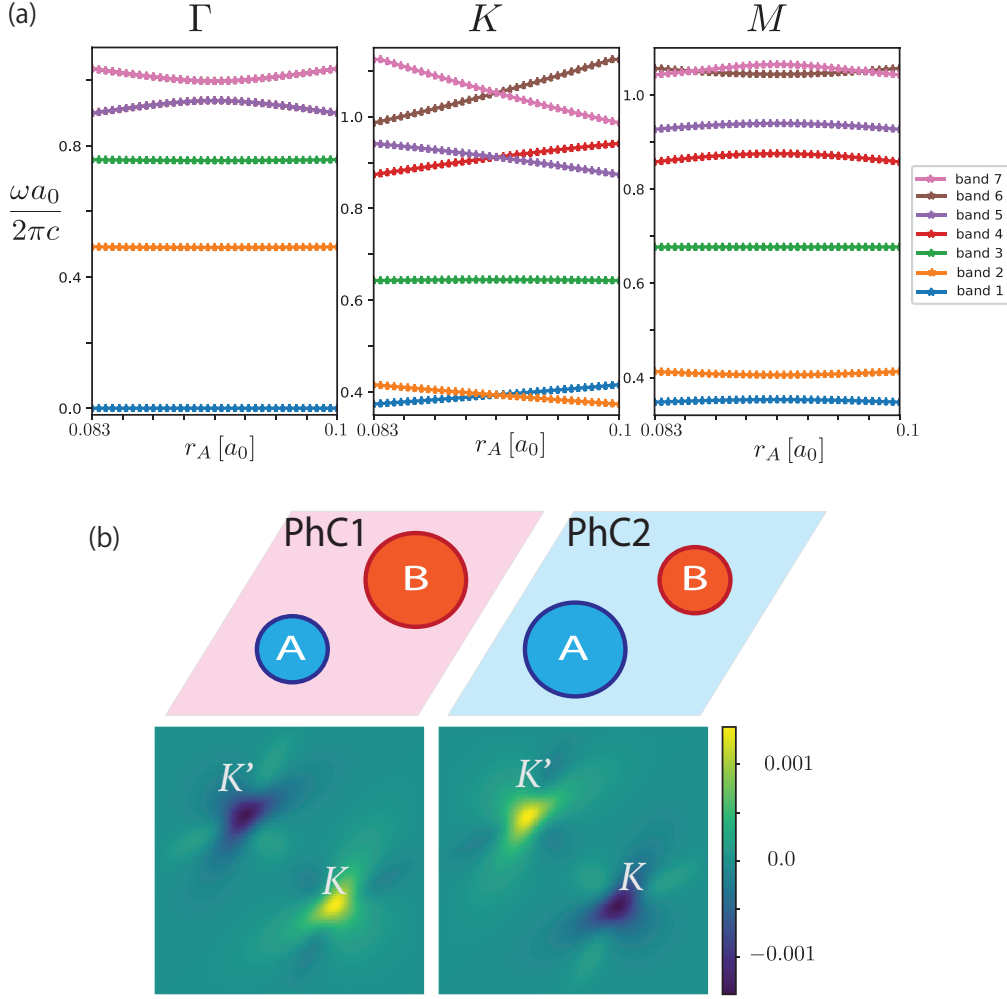}
\caption{(a) The radius size dependence of eigenfrequency for the seven
 lowest EM modes at high symmetric points, where the sum of $r_A$ and
 $r_B$ is kept constant, i.e., $r_A+r_B=0.183$. (b) Two inverted PhC
 structures and their corresponding Berry curvature distribution in the
 1st BZ. PhC2 is obtained by exchanging the position of two rods of
 PhC1.} 
\label{figure2}
\end{figure}

In Fig.~\ref{figure2}(a), the evolution of eigenfrequency for the
seven lowest EM modes is shown at high symmetric points $\Gamma$, $K$ and
$M$ in the 1st BZ, where the rods radii $r_A$ and $r_B$ are varied. 
However, it is assumed that the sum of $r_A$ and $r_B$ is kept constant,
i.e., $r_A+r_B=0.183 a_0$. 
It should be noted that the band inversions 
between 1st and 2nd, between 4th and 5th, and between 6th and 7th photonic
subbands happen at $K$ point when $r_A=r_B$. The band inversion indicates
the topological phase transition of bulk EM waves. 

From Fig.~\ref{figure2}(a), we can also confirm that 
photonic band structure remains unchanged even if $r_A$ and $r_B$
are exchanged.
Because of this symmetric property, we consider two inverted PhC
structures as shown in Fig.~\ref{figure2}(b). 
Let us define PhC with $r_A< r_B$ as PhC1, and
PhC with $r_A> r_B$ as PhC2, respectively. 
These two PhC structures share the same photonic band structure given in
Fig.~\ref{figure1}(d). However, as can be seen above, the Berry
curvature distribution in momentum space becomes different between PhC1
and PhC2.

In the broken inversion symmetry systems, the topological transition can be characterized
by the Berry curvature distribution in $\boldsymbol{k}$-space, i.e., 
\begin{equation}
    \boldsymbol{\Omega}_n\left(\boldsymbol{k}\right) = \boldsymbol{\nabla_k} \times \boldsymbol{A}_n\left(\boldsymbol{k}\right),
\end{equation} 
where $\boldsymbol{A}_n\left(\boldsymbol{k}\right) = i
\braket{u_{n,\boldsymbol{k}}|\nabla_{\boldsymbol{k}}|u_{n,\boldsymbol{k}}}$
is Berry connection and $|u_{n,\boldsymbol{k}}\rangle$ is periodic Bloch function of electric field. 
Owing to the broken inversion symmetry, the Berry curvature in the 1st BZ has
opposite sign between $K$ and $K^\prime$ points as indicated in
Fig.~\ref{figure2}(b). For PhC1, Berry curvature distribution around $K$ point have
positive value, however, that around $K^\prime$ point have the same magnitude but
negative sign, i.e., $\bm{\Omega}_n(\bm{K})=-\bm{\Omega}_n(\bm{K^\prime})$. 
In further, PhC2 has reversed Berry curvature distribution compared with
PhC1, i.e., $\bm{\Omega}_n(\bm{k})|_{PhC1}=-\bm{\Omega}_n(\bm{k})|_{PhC2}$.
Since
the 1st BZ is closed, the integration of Berry curvature over the 1st BZ is
quantized in the unit of $2\pi$~\cite{Dario}. Therefore, Chern number for a 2D closed surface is defined as 
\begin{equation}
    C_n = \frac{1}{2\pi}\oiint_{BZ} \boldsymbol{\Omega}_n\left(\boldsymbol{k}\right)d^{2}k.
\end{equation}
As can be seen in lower panel of Fig.~\ref{figure2}(b), 
$\boldsymbol{\Omega}_n\left(\boldsymbol{k}\right)$ has large values at
$K$ and $K^\prime$ points which give the large weight for this
integration. Since Berry curvatures at $K$ and $K^\prime$ points have opposite sign,
Chern number is identically 0. However, if we focus on one of two Dirac points, Chern number is not zero. Thus, we shall introduce the valley
Chern number as 
\begin{equation}
    C^{\mu}_n = \frac{1}{2\pi}\oiint_{S_\mu}
     \boldsymbol{\Omega}_n\left(\boldsymbol{k}\right)d^{2}k,
\end{equation}
where $\mu=K, K^\prime$. The integration area $S_\mu$
indicates the $\frac{1}{2}$ of 1st BZ which includes only $K$ or $K^\prime$ point.

\section{Topological interface states}
In this section, we discuss the localization of EM waves at 1D
zigzag interface between PhC1 and PhC2. The total Chern number of the 1st
band for both PhC1 and PhC2 is $0$, because each of PhCs
preserves time reversal symmetry~\cite{Raghu2008}. However, the 
valley Chern number at each valley has non-zero value, which is similar
to ref.~\cite{Chen2017}. In details, the valley Chern number for PhC1 is
$C_1^{K}|_{\rm PhC1} = 1/2$, 
$C_1^{K^\prime}|_{\rm PhC1} = -1/2$ and for PhC2 is $C_1^{K}|_{\rm PhC2} = -1/2$,
$C_1^{K^\prime}|_{\rm PhC2} = 1/2$.
\begin{figure}[ht!]
\centering\includegraphics[width=\textwidth]{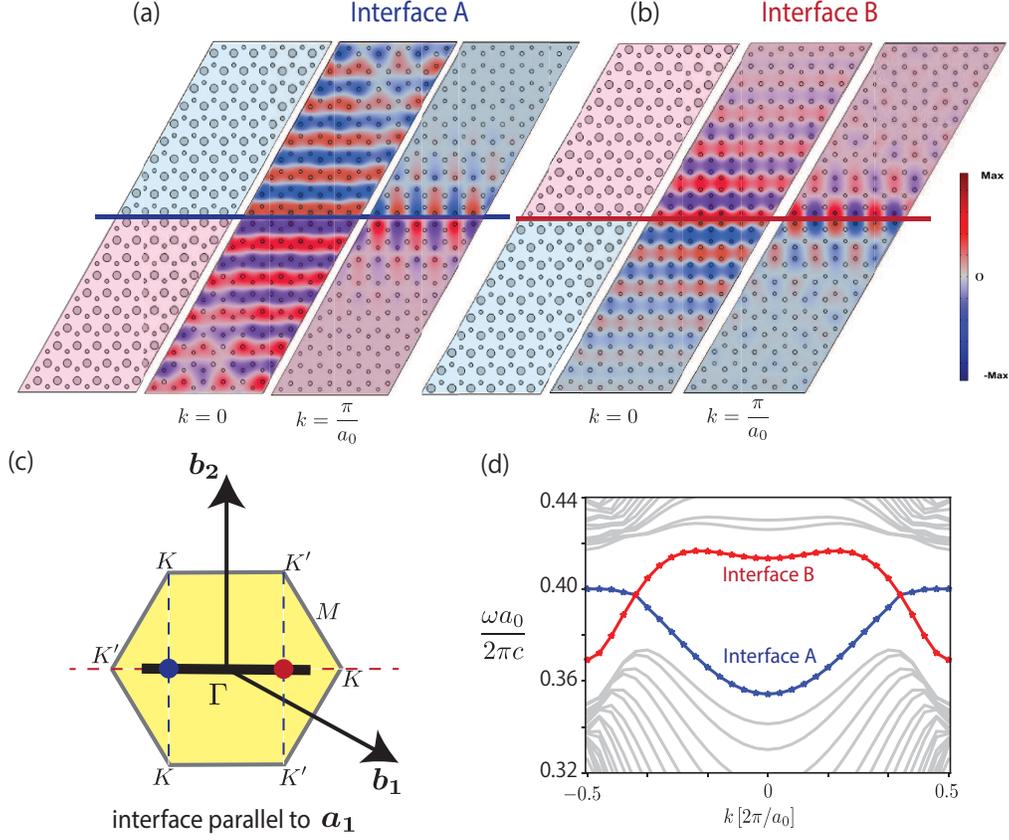}
\caption{The schematic of zigzag interface type A (a) and
interface type B (b) between PhC1 and PhC2 (left panels), which has the translational invariance along both $\bm{a_1}$ and $\bm{a_2}$ directions. The corresponding field profile of each interface structure at $k=0$ (middle panels) and $k=\frac{\pi}{a_0}$ (right panels). Magenta regions indicate PhC1 and cyan
regions indicate PhC2.
(c) The corresponding 1st BZ (thick black line) for the interface
structure which is parallel to $\boldsymbol{a_1}$ direction.
Blue (red) dot indicates the projection of $K(K^\prime)$ points
on to the 1st BZ.
(d) Photonic band structure for the interface structure
near the 1st photonic band gap. Gray lines are bulk states.
The blue line is the interface states for the interface A. The red line is the interface states for the
interface B. 
}  
\label{figure3} 
\end{figure}

According to the simple theoretical observations, at the interface
between PhC1 and PhC2, the valley Chern number for each 
$K$ points 
is quantized as: 
\begin{align}
&\Delta C^{K}=C_1^K|_{\rm PhC1}-C_2^K|_{\rm PhC2}=1/2-(-1/2)=1,\\
&\Delta C^{K^\prime}=C_1^{K^\prime}|_{\rm PhC1}-C_2^{K^\prime}|_{\rm PhC2}=-1/2-1/2=-1.
\end{align}
Thus, $K$ and $K^\prime$ valleys have quantized Chern number with opposite
sign, 
which leads to topological phase transition at the
interface between PhC1 and PhC2. As a consequence of topological phase
transition, highly localized EM states appear at the interface, i.e., 
topological interface (edge) states. Such topological interface EM states propagates
along the interface and quite robust against the scattering from
interface roughness and crystal defects.

In the left panels of Figs.~\ref{figure3}(a) and (b),
two different types of interface structure between PhC1 and PhC2 
are shown. We call them the interface A and B, respectively. 
For the interface A, PhC2 is placed upper
semi-half region, but PhC1 is placed lower semi-half region. 
In this case, larger dielectric cylinders come to at the interface.
For the interface B, the regions for PhC1 and PhC2 are inverted, 
then, smaller dielectric cylinders come to at the interface.
Here the interface is taken along the zigzag edge direction of 2D
honeycomb lattice, i.e., the interface has the translational invariance
along $\bm{a_1}$ direction. 
To distinguish between interfaces A and B, we shall define the valley
Chern number for interface as follows:
\begin{align}
& C^{K}_A\coloneqq C_1^K|_{\rm PhC2}-C_2^K|_{\rm PhC1}=-1/2-1/2=-1,\\
& C^{K^\prime}_A\coloneqq C_1^{K^\prime}|_{\rm PhC2}-C_2^{K^\prime}|_{\rm PhC1}=1/2-(-1/2)=1,\\
& C^{K}_B\coloneqq C_1^K|_{\rm PhC1}-C_2^K|_{\rm PhC2}=1/2-(-1/2)=1,\\
& C^{K^\prime}_B\coloneqq C_1^{K^\prime}|_{\rm PhC1}-C_2^{K^\prime}|_{\rm PhC2}=-1/2-1/2=-1.
\end{align}
Here, we have defined the valley Chern number of $\alpha(=K, K^\prime)$
valley for the interface $\beta(=A,B)$ as $C^\alpha_\beta$. 
It should be noted that the valley Chern numbers satisfy the following
relations:
\begin{align}
C^{\alpha}_A&=-C^{\alpha}_B,\\
C^{K}_\beta&=-C^{K^\prime}_\beta.
\label{eq.Chernrelation}
\end{align}
In the middle and right panels of Figs.~\ref{figure3}(a) and (b),
the spatial distribution of electric field is
shown at $k=0$ and $k=\pi/a_0$, respectively. 
Owing to the presence of finite valley Chern number, the localized
EM states are observed along the interfaces. 
Electric field
is highly localized at the domain wall between two PhCs at
$k=\pi/a_0$ and much more penetrate into the bulk at $k=0$.
It should be also noted that EM waves of interface states have opposite
group velocities near $K$ and $K^\prime$ points between interfaces A and B, which can be confirmed from
the photonic band structure shown in Fig.\ref{figure3}(d).

In Fig.~\ref{figure3}(c), the projected 1st BZ for interface
structures is denoted as a thick black line. The $K$ and $K^\prime$
points are projected to $-2\pi/3a_0$ and $2\pi/3a_0$, respectively. 
The photonic energy band structure for interfaces $A$ and $B$ is shown 
in Fig.~\ref{figure3}(d). The blue line is the interface states for the
interface A, while the red line is the interface states for 
interface B. 
Grey lines represent bulk states originated from 1st and 2nd subbands. 
It is clearly observed that topological interface states appear inside
photonic bandgap for bulk PhC as being observed before in \cite{Chen2017}. However, while only one type of interface structure were examined in \cite{Chen2017}, our results show two possible zigzag interface structures, which are created from PhC1 and PhC2.

\section{Topological corner states}
Recently, the localization of EM wave at 0D corners, so-called corner
states, 
as higher-order topology were examined in several PhC structures,
where the topological corner states are explained by
vectored Zak phase and gapped Wannier bands~\cite{Delplace2011, Liu2019,
Benalcazar2017}. Here we show that the localized EM corner states can
occur even in the honeycomb PhC with broken inversion symmetry. 
We will see that selection rule of corner state formation in the present
system, i.e. corner states appear for 60$^\circ$ corner, but absent for
120$^\circ$ corner. This selection rule can be elucidated as the
consequence of valley-valley interaction. 
\begin{figure}[ht!]
\centering\includegraphics[width=\textwidth]{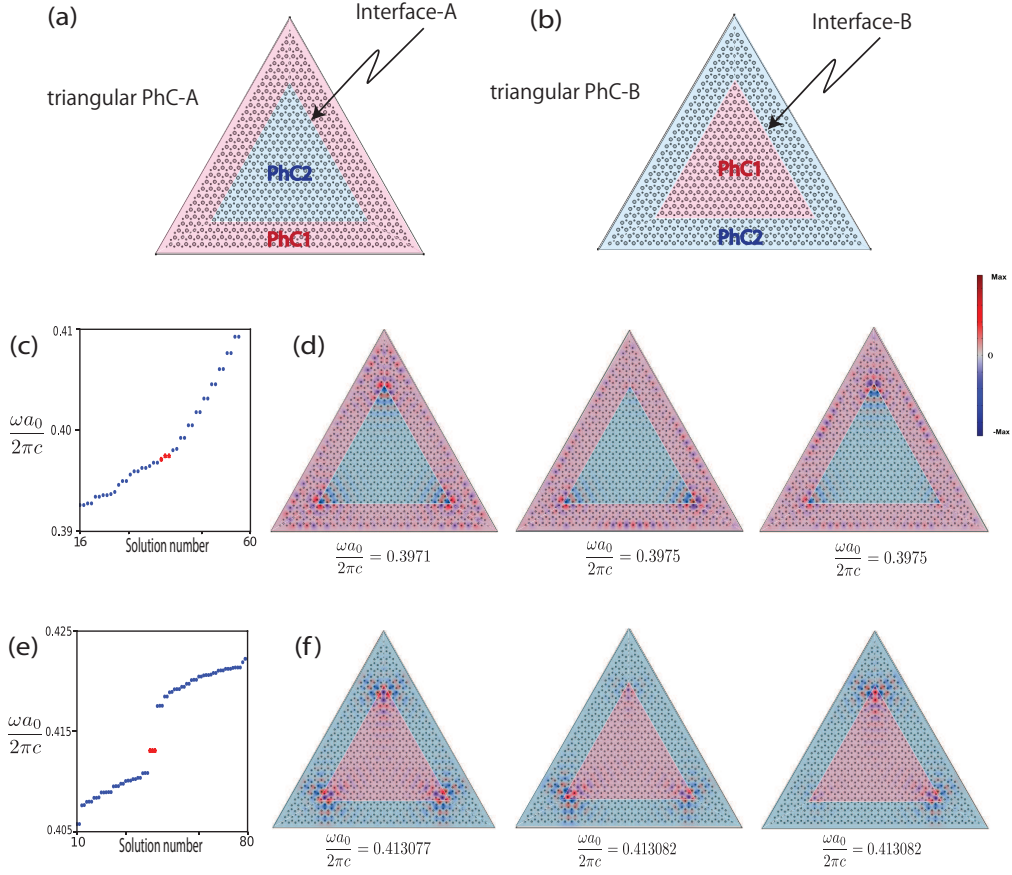}
\caption{(a) Structure of triangular PhC-A. The triangular PhC2 is
 embedded inside triangular PhC1, in which the interface structure
 becomes interface A. (b) Structure of triangular PhC-B. Compared with PhC-A,
 PhC1 and PhC2 are mutually replaced. The interface structure is
 interface B. 
(c) Photonic angular frequency spectrum of triangular PhC-A. The red
 dots indicate of eigenfrequencies for corner states. Three corner
 states appear.
(d) Electric field profile of three corner states for triangular PhC-A.
(e) Photonic angular frequency spectrum of triangular PhC-B. 
(f) Electric field profile of three corner states for triangular PhC-B. 
}  
\label{figure4}
\end{figure}

To observe corner states, we consider triangular PhC structure in which
another type triangular PhC is embedded. 
Figures~\ref{figure4} (a) and (b) show such triangular PhCs.
In Fig.~\ref{figure4}(a), triangular PhC2 is embedded in the triangular
PhC1. In this case, 
the interface between PhC1 and PhC2 becomes the
interface A, hereafter, we call triangular PhC-A.
Figure~\ref{figure4}(b) shows another type of triangular PhC, where PhC1 is embedded
in PhC2. In this case, 
the interface between PhC1 and PhC2 becomes the
interface B, so similarly we call triangular PhC-B hereafter.
Both of structures have three $60^\circ$ corners. 
To implement the numerical calculations, the perfect electric conductor
(PEC) boundary condition is imposed for outside triangular PhCs.
The structure contains $20$ lattice sites for inner triangular
PhC, and $32$ lattice sites for outer triangular PhC.

Figures~\ref{figure4}(c) shows the photonic angular frequency spectrum
of triangular PhC-A. The red dots indicate the eigenfrequencies of
three corner states.
Figures~\ref{figure4}(d) shows the field profile for each corner state.
Electric field of each corner state is localized at one of two sublattices at
the corners, then decays exponentially.

Similarly, Figure~\ref{figure4}(e) shows the photonic angular frequency
spectrum of triangular PhC-B, where the red dots indicate the corner
states. In this case, three corner states appear inside the photonic band gap,
i.e., well-separated from bulk and interface states.
Figure~\ref{figure4} (f) show the field profile for each corner states.

It can be seen that the corner states of triangular PhC-A are mixed with bulk and edges states. But the corner states of triangular PhC-B are isolated. This difference can be understood by inspecting the photonic band structure for the interface structures A and B as shown in Fig. 3 (d). The frequency range of interface A is completely overlapped with that of bulk and interface B. However, in the frequency range from 0.4 to 0.415, only interface B states are found. In the triangular structure, there are three outside edges with PEC boundary condition, so-called PEC edges. For triangular PhC-A, the PEC edges have the same structure as interface B. Therefore, in the gap of bulk photonic band, we find no frequency range of only interface A, resulting in the mixing the corner states with the bulk and PEC edges states. For triangular PhC-B, the PEC edges have the same structure as interface A. In the frequency range from 0.4 to 0.415, there are only the states of interface B. Therefore, the corner states of triangular PhC-B are isolated from bulk and PEC edge states.

The eigenfrequencies of three corner states in both cases are mutually close. But, one of
them is slightly lower than others. This is due to the interaction
between three corner states. This behavior can be understood by
employing simple three-sites tight-binding model. 
Let us represent the spacially localized electric field at three corners 
as $\phi_1$, $\phi_2$ and $\phi_3$, respectively. Then, we shall assume that
these three spatially localized electric fields
mutually interfere with the strength of $-\gamma$. 
Here, $\gamma$ is a parameter to represent the strength of interference
and minus sign indicates that the constructive interference makes energy lower.
So then, we can construct 
effective tight-binding Hamiltonian as
\begin{align}
 H=-\gamma\left(|\phi_1\rangle\langle \phi_2|+
|\phi_2\rangle\langle \phi_3|+
|\phi_3\rangle\langle \phi_1|
\right) + {\rm H.c.}.
\end{align}
This three-sites model gives three eigenvalues as 
$-2\gamma$, $\gamma$, $\gamma$, i.e., two of them are degenerate. The corresponding
eigenvectors are $(\phi_1,\phi_2,\phi_3)=(1,1,1)$, $(1,0,-1)$ and
$(-1,1,0)$, respectively. 
Thus, as shown in the Figs.~\ref{figure4}(e) and (g), three corner states have
almost same amplitudes at the corners, but in middle and right panels, one of corner
states is missing.

\begin{figure}[ht!]
\centering\includegraphics[width=0.8\textwidth]{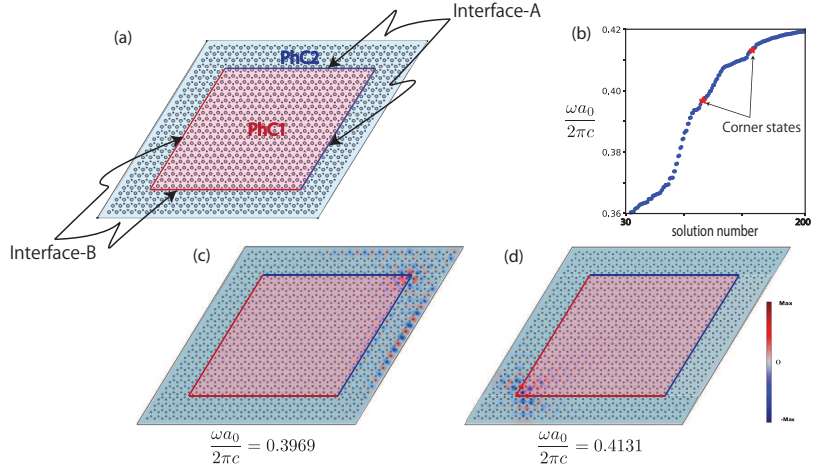}
\caption{(a) Rhombus structure of PhC1 and PhC2, consisting of both interface A (blue lines) and interface B (red lines). (b) The frequency spectrum for the rhombus structure. Red dots indicate the frequencies of $60^\circ$ corner states. The corresponding field profile of $60^\circ$ corner states which is form by interface A (c) and interface B (d).} 
\label{figure5}
\end{figure}

We also examined another structure, which includes both $60^\circ$
and $120^\circ$ corners as shown in Fig.~\ref{figure5}(a). 
The blue and red lines denote the interface A and interface B.
The structure has the size of $28\times 28$ supercell of PhC2. Inside
the PhC2, PhC1 with the supercell of $20\times 20$ is embedded. 
PEC boundary condition has been imposed for the outer surface of PhC2
for numerical calculation. 
Figure~\ref{figure5}(b) shows the photonic spectrum of the rhombus
structure. The red dots indicate frequencies for the states where
electric field is localized at the $60^\circ$ corners. The corner state
with lower frequency has the field profile shown in
Fig.~\ref{figure5}(c). This state is slightly mixed with the interface
states. The corner state with higher frequency has the field profile
shown in Fig.~\ref{figure5}(d). This corner state is isolated from other
interface and bulk states. These results matches with the cases of triangular structures. 
As for the $120^\circ$ corner, localized corner states are absent.

\begin{figure}[ht!]
\centering\includegraphics[width=1.0\textwidth]{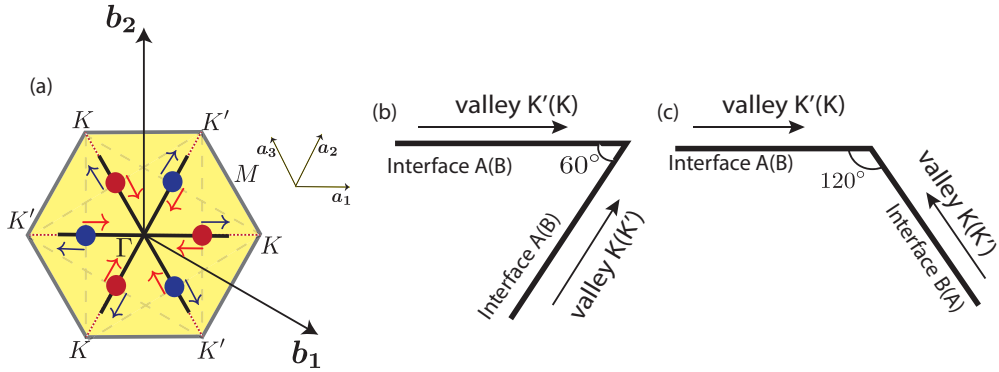}
\caption{(a) The 1st BZ (thick black lines) for the interface structures
 which are parallel to $\bm{a_1}$, $\bm{a_2}$ and $\bm{a_3}$
 directions. The blue and red circles indicate the projection of $K$ and
 $K^\prime$ points on to the 1st BZ. Blue and red arrows indicate the
 directions of group velocity at each valley for interface A and
 interface B, respectively. 
(b) Schematic structure of 60$^\circ$
 corner, which connect two interfaces with same type.
(C) Schematic structure of 120$^\circ$
 corner, which connect two interfaces with different type.
}
\label{figure6}
\end{figure}

Now we elucidate why the topological corner states appear at 60$^\circ$
corner, but absent in 120$^\circ$ corner. In previous section, we have
considered the zigzag interfaces which are parallel to
$\boldsymbol{a_1}$ direction shown in Figs.~\ref{figure3}(a) and (b). In
similar manner, we can also build the zigzag interfaces which are
parallel to either $\boldsymbol{a_2}$ or $\bf{a_3}$ direction. The black
bold lines in Fig.~\ref{figure6}(a) indicate the 1D 1st BZ of
projected band structure for zigzag interface in $\boldsymbol{a_1}$,
$\boldsymbol{a_2}$ and $\bf{a_3}$ directions. The blue and red arrows
indicate group velocity direction at each valley of interface A and
interface B, respectively. The group velocity $v_g$ is given by $v_g =
\nabla_{\bm{k}}\omega({\bm{k}})$.
Figure~\ref{figure6} (b) is the schematic of
$60^\circ$ corner, which includes two interfaces A or two
interfaces B. 
EM waves of interface states propagating toward the corner belong to
different valley.  
Thus, 60$^\circ$ corners always connect two interfaces with different
valley Chern number, i.e., the formation of corner states at $60^\circ$
corners. On the other
hand, Figure~\ref{figure6} (c) is the schematic of $120^\circ$ corner,
which includes both interface A and interface B. In 
this case, EM waves propagating toward the corner belong to 
the mutually opposite valley.
However,
since the valley Chern number are opposite between interfaces
A and B (see Eq.(\ref{eq.Chernrelation})), 
$120^\circ$ corners connect two interfaces with same valley Chern number, i.e., no formation
of corner states at  $120^\circ$ corner.
Thus, the sign flip of valley Chern numbers is necessary condition of
corner state formation.

The $60^\circ$ corners connect two interfaces with the opposite valley
Chern number, but $120^\circ$ corners connect two interfaces
with the same valley Chern number. Thus, the angle-dependent
formation of corner states are attributed to the valley Chern number.
To describe this, we shall introduce
the pseudo-spin for valley Chern number. 
In the following, we assume that the valley Chern number $1$ 
is pseudo-spin up and the valley Chern number
$-1$ is pseudo-spin down.
The two interfaces of $60^\circ$ corner
will have the interaction between two different valleys, 
which can be understood by the coupling of pseudo-spin up and
pseudo-spin down. 
However, for the $120^\circ$ corner, two interfaces show the same valley 
Chern number. Therefore, 
there is no interaction between two interfaces of $120^\circ$ corner.

To formulate this argument for the corner states, we introduce the pseudo-spin
states as 
$\ket{\uparrow} = (1,0)^T$, 
$\ket{\downarrow} = (0,1)^T$ to represent valley Chern number $C=1$ and
$-1$, respectively. Here we have omitted the valley and interface
indices, because they are not relevant in the following. 
And, $(\cdots)^T$ indicates the transpose of vector.
Then we construct the general form of Hamiltonian, $H_{\rm corner}$, to represent the
formation of corner state as the interaction between two states with
different valley Chern number, i.e., 
\begin{equation}
    H_{\rm corner} = \varepsilon_0  I + \nu \boldsymbol{k} \sigma_{z}  + m \sigma_{x} + m \sigma_{y},
\end{equation}
where $\varepsilon_0$ and $\nu$ are the
eigenfrequency 
and group velocity at $K$ and $K^\prime$ points, respectively.
Here, $I$ is unit matrix, $\sigma_x$, $\sigma_y$ and $\sigma_z$
are Pauli matrices, $m$ is real value.
Note that the group velocities have mutually opposite directions at $K$ and $K^\prime$ points.
The first two terms of $H_{\rm corner}$ give the
diagonal part of Hamiltonian, which represents energy for each specific
interface. Since valley Chern number at $K$ and $K^\prime$ points are finite,
we represent the interaction between the different 
valleys of two interfaces as the third and the forth terms. The real
value $m$ shows the strength for the interaction, which depends on the
structures and materials. The $60^\circ$ corners have finite interaction
between two interfaces, leading to the scattering between pseudo-spin
states. On the other hand, for 
$120^\circ$ corners, $m$ should be zero, because there is no interaction
between $K$ and $K^\prime$ points, i.e., conservation of pseudo-spin
state. 

Finally we shall mention about other corner angles such as $30^\circ$, $90^\circ$,
$150^\circ$. These corners are constructed as the combination of
zigzag and armchair edges. Thus, the corners always contain an armchair
edge. As for armchair edges, $K$ and $K^\prime$ points come to $k=0$ in
the projected 1D 1st BZ, i.e., $\Delta C^K+\Delta C^{K^\prime}=0$. 
Thus, the
valley Chern number becomes identically zero. 
Because the sign change of valley Chern number are not involved at
these corners, no corner state appears at $30^\circ$, $90^\circ$ and $150^\circ$ corners. 
It is noted that this observation is similar to the graphene corner
states ~\cite{liu2020,shimomura}, but where the formation mechanism of
corner states relies on Zak phase, not valley Chern number. 
 
\section{Conclusion}
In this paper, we have numerically studied the EM states in the
honeycomb PhC structure using plane wave expansion and finite element
methods. The system is composed of dielectric cylinders which are
periodically arrayed on the 
honeycomb lattice. Owing to the honeycomb lattice symmetry, the unit
cell contains two non-equivalent dielectric cylinders with the radii
$r_A$ and $r_B$. 

In case of $r_A = r_B$, the system possesses
inversion symmetry and becomes photonic analogue of graphene, i.e., Dirac cones at $K$
and $K^\prime$ points in the 1st BZ. 
However, if two cylinders have different diameters, i.e., $r_A\neq r_B$, the inversion symmetry is broken,
resulting in the lift of Dirac points in 
the photonic band structure, i.e., photonic bandgap opening. 
In this situation, 
$K$ and $K^\prime$ points have finite Berry curvature but with opposite
signs. Since the system respects with time-reversal symmetry, total Chern number is always zero.
However, $K$ and $K^\prime$ possess finite valley Chern number with
mutually opposite sign. 

The hidden topological properties of valley Chern number can be
extracted by considering the interface between two different PhCs,
where one side is PhC with $r_A > r_B$, the other side is $r_A<r_B$. 
Along the interface, we have found the formation of the interface states
owing to the finite valley Chern number.

It is found that 60$^\circ$ corner structure can
induce the localized corner states as a consequence of sign flip of
valley Chern number at corners.
However, such corner states are absent for 120$^\circ$,
because no sign flip of valley Chern number is involved. 
It is also pointed out the corners with $30^\circ$, $90^\circ$ and $150^\circ$
also do not possess the corner states, because these corner involves
armchair edges where total valley Chern number identically zero.

Our results suggest the topological design of photonic crystal on the
basis of valley degree of freedom.
The topological interface states can be useful for 
the high efficient optical communication technology, and the corner states
will be useful for the confinement of electromagnetic waves.

\section*{Acknowledgments}
K.W. acknowledges the financial support by JSPS KAKENHI (Grant
No. 21H01019 and JP18H01154), and JST
CREST (Grant No. JPMJCR19T1). F. L. acknowledges the financial support by Research Starting Funding of Ningbo University and NSFC Grant No. 12074205.

\section*{Disclosures}

The authors declare no conflicts of interest.

\bibliography{references}

\end{document}